\documentclass[aps,prb,twocolumn,showpacs,showkeys,preprintnumbers,groupedaddress,amsmath,amssymb]{revtex4-2}

\usepackage{graphicx}
\usepackage{dcolumn}
\usepackage{bm}
\usepackage{color}
\usepackage{amsmath}
\usepackage{hyperref}

\begin{document}


\preprint{}


\title{Crossover from Cooper-pair hopping to single-electron hopping in Pb$_x$(TiO$_2$)$_{1-x}$ granular films}


\author{Zhi-Hao He}
\affiliation{Tianjin Key Laboratory of Low Dimensional Materials Physics and
Preparing Technology, Department of Physics, Tianjin University, Tianjin 300354,
China}

\author{Zi-Yan Han}
\affiliation{Tianjin Key Laboratory of Low Dimensional Materials Physics and
Preparing Technology, Department of Physics, Tianjin University, Tianjin 300354,
China}

\author{Kuang-Hong Gao}
\affiliation{Tianjin Key Laboratory of Low Dimensional Materials Physics and
Preparing Technology, Department of Physics, Tianjin University, Tianjin 300354,
China}

\author{Zi-Wu Wang}
\affiliation{Tianjin Key Laboratory of Low Dimensional Materials Physics and
Preparing Technology, Department of Physics, Tianjin University, Tianjin 300354,
China}

\author{Zhi-Qing Li}
\email[Corresponding author, e-mail: ]{zhiqingli@tju.edu.cn}
\affiliation{Tianjin Key Laboratory of Low Dimensional Materials Physics and
Preparing Technology, Department of Physics, Tianjin University, Tianjin 300354,
China}



\date{\today}

\begin{abstract}
The electrical transport properties of Pb$_x$(TiO$_2$)$_{1-x}$ ($x$ being the Pb volume fraction and ranging from $\sim$0.45 to $\sim$0.69) granular films are investigated experimentally. The charging energy of the Pb granules is reduced to less than the superconducting gap of Pb granules for the low temperature insulating films by using high-$k$ dielectric TiO$_2$ as the insulating matrix. For the insulating films in the vicinity of the superconductor-insulator transition, Cooper-pair hopping governs the low-temperature hopping transport. For these films, the low-temperature magnetoresistance is positive at low field and the resistivity vs temperature for Cooper-pair hopping obeys an Efros-Shklovsii-type variable-range-hopping law. A crossover from Cooper-pair-dominated hopping to single-electron-dominated hopping is observed with decreasing $x$. The emergence of single-electron-dominated hopping in the more insulating films is due to the causation that the intergrain Josephson coupling becomes too weak for Cooper pairs to hop between adjacent superconducting Pb granules.
\end{abstract}


\maketitle


\section{Introduction}\label{SecI}
Disordered superconductors have attracted great attention over the past few decades since a range of novel phenomena, such as superconductor-insulator transition in two-dimensional superconducting films~\cite{no1,no2,no3,no4,no5,no6,add1,add2}, anomalous metallic state~\cite{no7,no8,no9,no10,no11,no12,add4,add5}, macroscopic quantum tunneling effect~\cite{no13,no14,no15,no16}, and disorder-induced inhomogeneity (emergence of superconducting islands or puddles)~\cite{no17,no18,no19,no20}, are related to the interplay between the superconductivity and localization. Among the disordered superconductors, granular superconductor is an important experimentally accessible model system, which offer a unique testing ground for studying tunable combined effects of disorder, Coulomb interactions, and superconductivity~\cite{no21}. The knowledge on these effects is important in understanding the properties of high-temperature superconductors (the disorder-induced granularity was evidenced in high-temperature superconductors~\cite{no22,no23,no24,no25}) as well as in understanding the behaviors of devices related to Josephson junctions~\cite{no26,no27,no28,no28,no29,no30,no31}. In granular superconductors, the ground state of the system is ultimately determined by the competition of Josephson coupling energy $E_J$ and charging energy $E_c$. The granular superconducting arrays with $E_J < E_c$ are insulators, while those with $E_J > E_c$ are superconductors at low temperature~\cite{no21}. In the insulating regime and below the local superconducting transition temperature $T_c^L$, Cooper pairs are localized on the granules with sufficiently large size by Coulomb repulsion. Thus an interesting question is whether the electric transport below $T_c^L$ is dominated by the hopping of Cooper pairs or single electrons at the insulating regime.


Theoretically, it has been proposed that the current through a superconducting grain is sensitive to the relation between the superconducting gap $\Delta$ and the charging energy $E_c$. For $E_c > \Delta$, the transport is governed by single-electron hopping, while the hopping transport is mediated by Cooper pairs in the opposite case $E_c < \Delta$~\cite{no32,no33,no34}. When a magnetic field is applied, the single-electron-hopping could result in a negative magnetoresistance and the Cooper-pair hopping should lead to a positive magnetoresistance due to the suppression of the order parameter~\cite{no4}. In addition, Lopatin and Vinokur further proposed that the temperature dependence of conductivity would obey the Efros-Shklovsii (ES) type variable-range-hopping (VRH) law in the Cooper pair hopping regime~\cite{no34}. In the insulating regime of granular superconductors, the single-electron-hopping conduction is widely observed in experiments~\cite{no35,no36,no37,no38}. However, the Cooper pair hopping transport has only been reported in a limited number of low dimensional granular systems, including one-dimensional Josephson junction chains~\cite{no29}, two-dimensional (2D) Josephson junction arrays~\cite{no26}, and 2D InO$_x$ and In-InO$_x$ granular films~\cite{no39,no40}. In 2D Josephson junction arrays, the temperature dependence of the Cooper pair hopping conductance was found to obey the Arrhenius form equation~\cite{no26}, while an ES-type-VRH conductance was observed in the 2D InO$_x$ and In/InO$_x$ granular films~\cite{no39,no40}.  Thus, more comprehensive investigations on the transport properties of Cooper pair hopping in granular superconductor are desirable. On the other hand, the physical properties in three-dimensional (3D) systems may differ from those in low-dimensional systems. It is therefore nontrivial to check whether the Cooper pair hopping, as well as the ES-type-VRH conduction of Cooper pair, could be realized in 3D granular superconductors.

To satisfy the condition for Cooper-pair hopping, $E_c < \Delta$, one needs to choose a suitable superconductor-insulator granular system with relative small $E_c$ and large $\Delta$. For a superconductor-insulator granular composite, the charging energy of a grain can be estimated via $E_c\simeq e^2/(4\pi\epsilon_0 \tilde{\kappa}a)$~\cite{no32}, where $e$ is the electronic charge, $\epsilon_0$ is the permittivity of free space, $\tilde{\kappa}$ is the effective dielectric constant and proportional to the permittivity of the insulator constituent $\epsilon_r$, and $a$ is the mean grain size. Therefore, an effective way of reducing the charging energy is to select a high-$k$ dielectric as the insulating matrix of the granular composites.

In the present paper, we successfully realized the Cooper-pair hopping transport and observed the ES-type-VRH conductivity~\cite{ES-Hop1,ES-Hop2,ES-Hop3,no44} of Cooper pairs in the insulating regime (near the superconductor-insulator transition) of Pb$_x$(TiO$_2$)$_{1-x}$ ($x$ being the Pb volume fraction) granular films. A crossover from Cooper-pair hopping to single-electron hopping is also observed as the films are extended to a more insulating region. It is noted that the room-temperature permittivity of TiO$_2$ is $\sim$10 ($\sim$20) times as large as that of Al$_2$O$_3$ (SiO$_2$)~\cite{no45}, and the zero-temperature superconducting gap of Pb is 1.37\,meV~\cite{no46}, being larger than the gaps of most elemental superconductors.

\begin{figure}[htp]
\begin{center}
\includegraphics[scale=1]{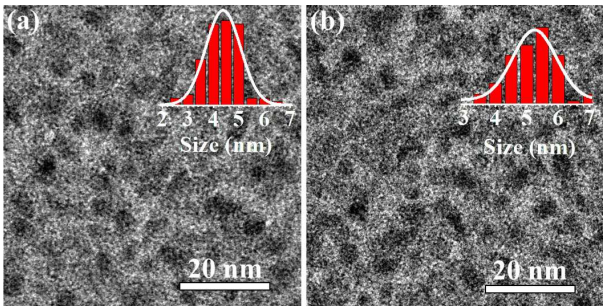}
\caption{\label{Fig-TEM} Bright-field TEM images for the Pb$_x$(TiO$_2$)$_{1-x}$ films with (a) $x\simeq 0.45$, and (b) $x\simeq 0.52$.}
\end{center}
\end{figure}

\section{Experimental Method}\label{SecEM}
The Pb$_x$(TiO$_2$)$_{1-x}$ films with $x$ ranging from $\sim$0.45 to $\sim$0.69 were deposited on glass substrates at room temperature by co-sputtering Pb and TiO$_2$ targets in Ar atmosphere. The details of the deposition procedures are similar to those described in Ref.~\cite{no38}. The thicknesses of the films ($\sim$400 to $\sim$600\,nm) were measured using a surface profiler (Dektak, 6M). The Pb volume fraction $x$ in each film was obtained from the energy-dispersive x-ray spectroscopy analysis (EDS; EDAX, model Apollo X). The microstructure of the films was characterized by transmission electron microscopy (TEM; Tecnai G2 F20). The resistance as well as the current vs voltage were measured using the standard four probe method and the temperature and magnetic field environments were provided by a physical property measurement system (PPMS-6000, Quantum Design). For films with large $x$, both the current source and the voltmeter were provided by the model 6000 PPMS controller, while for the small-$x$ films, a Keithley 236 and a Keithley 2182A were used as current source and voltmeter, respectively. The narrow rectangle shape films (1\,mm$\times$10\,mm), defined by mechanical masks, were used for the transport measurement.

\section{Results and discussion}
Figure~\ref{Fig-TEM}(a) and \ref{Fig-TEM}(b) show the bright-field TEM images for two representative films with $x\simeq 0.45$ and 0.52, respectively. The dark and bright regions represent Pb granules and TiO$_2$ matrix, respectively.  Both the electron diffraction and x-ray diffraction results (not shown) indicate that TiO$_2$ is amorphous. Thus the films are granular composites in which the Pb granules are embedded in the amorphous TiO$_2$ matrix. The mean size of Pb granules, $a$, obtained
by taking into account $\sim$80 grains for each film, decreases with decreasing $x$, e.g., the mean grain size is $\sim$6.1\,nm for the $x \simeq 0.69$ film, decreases to $\sim$5.1\,nm when $x\simeq 0.52$, and further decreases to $\sim$4.3\,nm as $x$ is reduced to 0.45. Since the thickness of the film is at least 50 times larger than the mean size of Pb granules, the Pb granules form three-dimensional granular arrays.

\begin{figure}
\begin{center}
\includegraphics[scale=0.9]{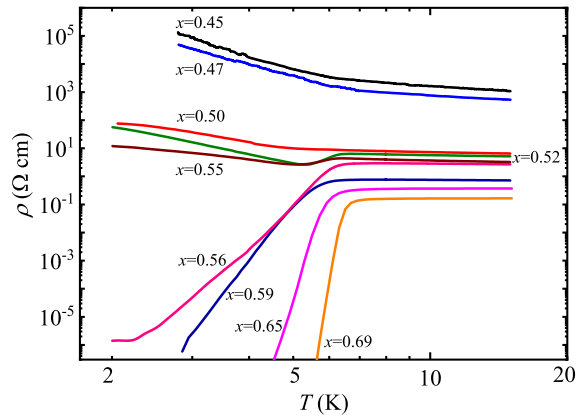}
\caption{Resistivity as a function of temperature for the Pb$_x$(TiO$_2$)$_{1-x}$ films ($0.45\lesssim x \leq 0.69$) at zero magnetic field.}\label{FigR-T-All}
\end{center}
\end{figure}

Figure~\ref{FigR-T-All} shows the temperature dependence of resistivity for the Pb$_x$(TiO$_2$)$_{1-x}$ films from $\sim$15 down to $\sim$2\,K. Clearly, global superconductivity appears in the films with $x\gtrsim 0.59$ at low temperatures. For the $x\simeq 0.56$ film, the resistivity  slightly increases with decreasing temperature initially, then sharply decreases about 6 orders of magnitude from $\sim$7.0 to $\sim$2.4\,K, and finally tends to be saturated with further decreasing temperature. Similar dissipation effect of $\rho(T)$ was also observed in Pb$_x$(SiO$_2$)$_{1-x}$ films with $0.60 \lesssim x \lesssim 0.72$ and have been explained by a model that includes both thermally activated phase slips and quantum phase slips~\cite{no38}. For the $x\simeq 0.55$ and 0.52 films, the temperature dependence of resistivity exhibits a reentrant behavior, i.e., the resistivity increases with decreasing temperature, starts to decreases with decreasing temperature at $\sim$7.0\,K, reaches to its minimum, and then increases with further decreasing temperature. According to Parendo \emph{et al.}~\cite{no47}, in granular films of amorphous bismuth the drop in resistivity is due to the occurrence of superconductivity on the isolated grains below the superconducting transition temperature of the grain, while the increase in resistivity at low temperature is caused by the opening of the energy gap on the superconducting grains coupled by quasiparticle tunneling (hopping). Then if the charging energy $E_c$ is less than the superconducting gap $\Delta$, the Cooper pairs do not have to decompose into electrons, and could conduct directly among the grains via ES-VRH~\cite{no34}. (For the Pb$_x$(TiO$_2$)$_{1-x}$ films, which mechanism makes the resistivity rise at low temperature will be amply studied below.) For the $x\lesssim 0.50$ films, the resistivity increases with decreasing temperature and increases more rapidly below $\sim$7.0\,K. We will focus our attention on the four films with $x\lesssim 0.52$ in the following discussion.

\begin{table}
\caption{\label{TableI} Relevant parameters for the Pb$_x$(TiO$_2$)$_{1-x}$ films with different $x$. Here $x$ is volume fraction of Pb, $t$ is the film thickness, $a$ is mean diameter of Pb granules, $\rho_N$ is the zero-field resistivity at 15 K, $E_{c}$ is the average charging energy, $E_J$ is the average Josephson energy between neighbor grains, and $\xi^{\rm {CPH}}$ is the Cooper-pair localization length for the ES-VRH hopping, $T_{0}$ is the characteristic temperature in Eq.(1).}
\begin{ruledtabular}
\begin{center}
\begin{tabular}{cccccccccccc}
 $x$   & $t$  & $a$   & $\rho_N$        & $E_c$    & $E_J$   & $T_0$   & $\xi^{\rm{CPH}}$  \\
       & (nm) & (nm)  & ($\Omega$\,cm)  &   (meV)  &  (neV)  & (K)     & (nm)   \\  \hline
0.52  &  562  & 5.1   & 5.18            & 0.64     &  435       & 0.9     &  0.9    \\
0.50  &  501  & 4.8   & 6.51            & 0.72     &  326       & 1.7     &  1.7     \\
0.47  &  376  & 4.5   & 535             & 0.83     &  3.7       &         &          \\
0.45  &  405  & 4.3   & 1080            & 0.91     &  1.8       &         &       \\
\end{tabular}
\end{center}
\end{ruledtabular}
\end{table}

\begin{figure}
\begin{center}
\includegraphics[scale=1.1]{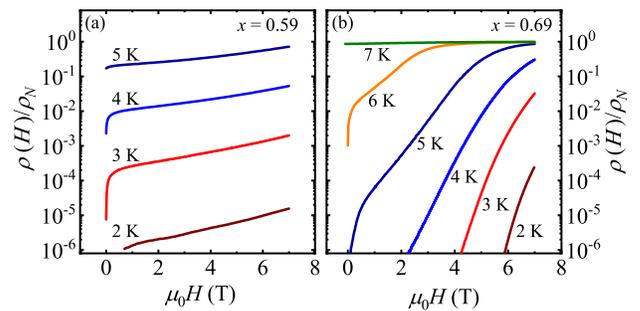}
\caption{\label{FigMRS} The normalized resistivity $\rho(H)/\rho_N$ vs magnetic field at different temperatures for (a) $x\simeq 0.59$ and (b) $x\simeq0.69$ Pb$_x$(TiO$_2$)$_{1-x}$ films. Here the normal state resistivity $\rho_N$ is taken as the resistivity at 15\,K.}
\end{center}
\end{figure}

Before presenting the hopping transport results for those $x\lesssim 0.52$ films, we first briefly discuss the low-temperature transport properties of the films with global superconductivity ($x\gtrsim 0.59$). Figure~\ref{FigMRS}(a) and \ref{FigMRS}(b) show the temperature dependence of the normalized resistivity $\rho/\rho_N$ at different temperatures for the $x\simeq 0.59$ and 0.69 films, respectively. Here the values of $\rho_N$ were taken as the resistivity at 15\,K for each film. Clearly, the resistivity at each measured temperature increases with increasing field, revealing typical characteristics of a type-II superconductor. For type-II superconductors, the upper critical field $H_{c2}$ for superconducting transition is  generally defined as the field at which the resistivity reaches to $\rho_N/2$. From Fig.~\ref{FigMRS}(a), one can see that the value of $H_{c2}$ for the $x\simeq 0.59$ film is much greater than 7.0\,T at 4.0\,K, while its value decreases to $\sim$5.6\,T at 5.0\,K. For the $x\simeq 0.69$ film, the values of $H_{c2}$ are greater than 7.0\,T below $\sim$4.0\,K and decrease to 5.5 and 2.7\,T at 5.0 and 6.0\,K, respectively. At a fixed temperature, $H_{c2}$ increases with decreasing $x$ for films with $x$ ranging from 0.59 to 0.69, which is quite similar to that for the Pb$_x$(SiO$_2$)$_{1-x}$ films with global superconductivity~\cite{no38}.  For our Pb$_x$(TiO$_2$)$_{1-x}$ films, the percolation threshold $x_c$ for metal to insulator transition is estimated to be $x_c \simeq 0.55$. Thus, in the films with global superconductivity there are many conducting paths composed of geometrically connected Pb granules. Then the charge carriers are mainly transported through those paths and do not need to hop between Pb granules in the conducting process. The Cooper pair hopping or single-electron hopping conduction, therefore, cannot be observed in these films.

\begin{figure}
\begin{center}
\includegraphics[scale=1.08]{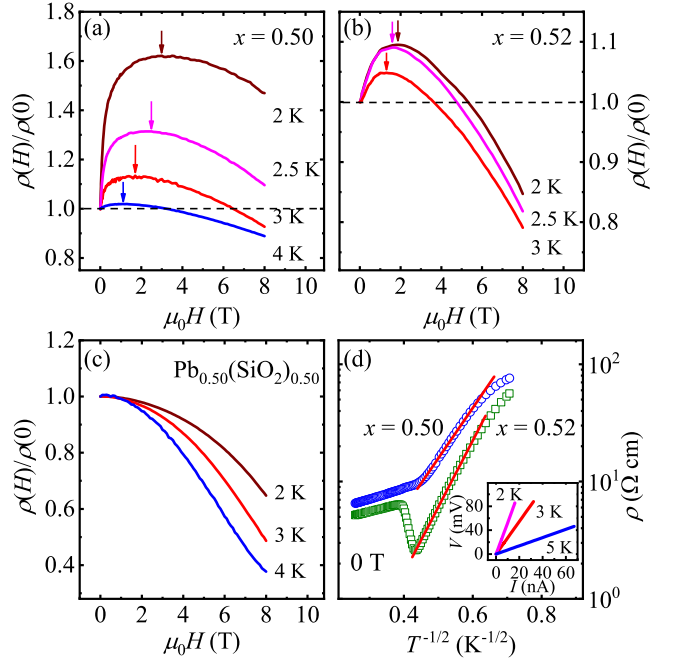}
\caption{\label{FigR-H1} The normalized resistivity $\rho(H)/\rho(0)$ vs magnetic field at different temperatures for (a) $x\simeq 0.50$ and (b) $x\simeq0.52$ Pb$_x$(TiO$_2$)$_{1-x}$ films, and (c) Pb$_{0.50}$(SiO$_2$)$_{0.50}$ film. The field  $H_{\rm P}(T)$ at which the resistivity reaches the maximum is indicated by the arrow. (d) Resistivity vs $1/T^{1/2}$ for the $x\simeq0.50$ and $x\simeq0.52$ Pb$_x$(TiO$_2$)$_{1-x}$ films at zero field. The inset shows the zero-field $I$-$V$ curves at different temperatures for the $x\simeq0.50$ Pb$_x$(TiO$_2$)$_{1-x}$ film.}
\end{center}
\end{figure}

Figure~\ref{FigR-H1}(a) and \ref{FigR-H1}(b) show the normalized resistivity $\rho(H)/\rho(0)$ as a function of magnetic field $\mu_0 H$ (with $\mu_0$ being the permeability of free space) at different temperatures for the $x\simeq 0.52$ and 0.50 films, where $\rho(H)$ and $\rho(0)$ are the resistivities at $H$ and zero field, respectively. Below $\sim$4\,K, the resistivity increases with increasing field at low field, peaks at $H_{\rm P}(T)$, then decreases with further increasing field above $H_{\rm P}(T)$, and finally tends to be less than 1.0 at high field regime (i.e., the magnetoresistance tends to be negative under high field). The field $H_{\rm P}(T)$ is sample dependent and decreases with increasing temperature for each film. For a single Pb grain, the critical size for existing superconductivity is $\sim$2.0\,nm~\cite{no48}, thus the granules remain superconducting in our insulating Pb$_x$(TiO$_2$)$_{1-x}$ films. The behaviors of low-temperature magnetoresistances strongly suggest that the transport process is governed by Cooper-pair hopping at least in low field regime: the magnetic field can destroy the Josephson coupling between the superconducting grains and suppress the amplitude of superconducting order parameter in each individual grain; when the magnetic field is relative small, the order parameter amplitude is almost unchanged, while the intergrain Josephson coupling is gradually suppressed with increasing field~\cite{no4}, which reduces the hopping probability of Cooper pairs between the superconducting grains and thus leads to a positive magnetoresistance.

Along with increasing field, the charging energy of some Pb granules with small size will be larger than the energy gap $\Delta(H)$. (The evolution of $\Delta$ with field can be estimated via $\Delta(H)=\Delta(0)[1-(H/H_{c2})^{1.6}]^{3/2}$~\cite{no26}.) This opens the channels for single-electron hopping, and will slow down the increasing rate ($\mathrm{d}\rho/\mathrm{d}H$) of the resistivity. At $H_{\rm P}(T)$, the decreasing rate of the conductivity~\cite{note1} caused by the reduction in the concentration of Cooper pairs (RCCP) for hopping is equal to the increasing rate of the conductivity resulted from the growth in the concentration of single-electron (GCSE) for hopping, $\rho(H)$ reaches its maximum. Above $H_{\rm P}(T)$, the decreasing rate of the conductivity by RCCP is less than the increasing rate of the conductivity by GCSE, the resistivity will decreases with further increasing field. When the reduction of the conductivity by RCCP is as large as the increment of the conductivity by GCSE, the magnetoresistance will be zero [$\rho(H)=\rho(0)$]. Since $\Delta(H)$ increases with decreasing temperature, the magnetoresistance peak, as well as the field for $\rho(H)=\rho(0)$, will shift to a higher field with decreasing temperature, which is what we see in Fig.~\ref{FigR-H1}(a) and \ref{FigR-H1}(b). In fact, for the $x\simeq 0.52$ and 0.50 films, Cooper pair hopping still dominates the low-temperature transport process even at a relatively high field used in the measurements. For example, when a field with magnitude of 8.0\,T is applied, the increments (reductions) of the conductivities (resistivities) for the $x=0.50$ and 0.52 films are less than 11\% (10\%) and 28\% (22\%), respectively, at 3.0\,K, which indicates that hopping transport is still governed by Cooper pair hopping at this situation.

We evaluate the average charging energies of the insulating Pb$_x$(TiO$_2$)$_{1-x}$ films using $E_c=e^2/(4\pi\epsilon_0\tilde{\kappa}a)$. The effective dielectric constant can be estimated via $\tilde{\kappa}=\epsilon_r(a+s)/s$ with $s$ being the intergrain spacing between neighboring grains~\cite{no50}. Assuming the Pb particles form a simple cubic lattice, one can roughly obtain $s$ via $x=a^3/(a+s)^3$. Taking $\epsilon_r$ of TiO$_2$ as $\sim$86~\cite{no45}, we obtain the values of $E_c$ for the films in the dielectric regime and list them in Table I. Clearly, the charging energy increases with decreasing $x$, and satisfies the condition for Cooper-pair hopping ($E_c < \Delta_0$ with the zero-temperature gap $\Delta_0\simeq 1.37$\,meV for the Pb granules) for the $x\lesssim0.52$ films. For comparison, we give the low temperature $\rho(H)/\rho(0)$ vs $\mu_0H$ data in Fig.~\ref{FigR-H1}(c) for a Pb$_{0.50}$(SiO$_2$)$_{0.50}$ film with thickness of $\sim$1\,$\mu$m and fabricated previously~\cite{no38}. This film is in insulating regime, and its resistivities at 2.0, 3.0, and 4.0\,K are 1086, 987 and 722\,$\Omega$\,cm, respectively. For this film, taking $a$ and $\epsilon_r$ as 7.9 nm and 4.5~\cite{no50}, respectively, one can obtain $E_c\simeq$ 8.4 meV, which is much greater than the energy gap $\Delta_0$ of Pb granules. Inspection of Fig.~\ref{FigR-H1}(c) indicates that the magnetoresistance at each temperature below $\sim$4\,K is negative over the whole measured field range (from zero to 8.0\,T). In fact, for the insulating Pb$_{x}$(SiO$_2$)$_{1-x}$ films near the superconductor-insulator transiton, such as $x\simeq 0.53$ to 0.51, the low-temperature magnetoresistances are all negative~\cite{no38}. Thus, the positive magnetoresistance in Pb$_x$(TiO$_2$)$_{1-x}$ ($x\simeq 0.50$, 0.52, and 0.55) films and negative magnetoresistance in the Pb$_{x}$(SiO$_2$)$_{1-x}$ film strongly suggest that the physical quantities governing the charge types of hopping in the insulating regime of granular superconductors are $E_c$ and $\Delta_0$.

Figure~\ref{FigR-H1}(d) shows the logarithm of resistivity as a function of $T^{-1/2}$ for the $x\simeq 0.50$ and 0.52 films, as indicated. Inspection of this figure indicates that the logarithm of resistivity varies linearly with $T^{-1/2}$ from $\sim$4.7 to $\sim$2.7\,K for the two films, which suggests that ES-type of Cooper-pair-hopping mechanism governs the conduction process in this temperature regime. In the ES-type Cooper-pair-hopping process, the resistivity obeys~\cite{no34}
\begin{equation}\label{Eq-ES-hopping}
 \rho(T)=\rho_{01} \exp\left(\frac{T_{0}}{T}\right)^{1/2},
\end{equation}
with
\begin{equation}\label{Eq-T0}
T_{0}=\frac{b(2e)^2}{4k_{B}\pi\epsilon_0\widetilde{\kappa}\xi^{\rm CPH}},
\end{equation}
where $\rho_{01}$ is a prefactor, $b$ is a numerical factor of the order of unity, $k_B$ is the Boltzmann constant, and $\xi^{\rm CPH}=a/\ln(8\bar{E}/\pi\bar{g}\bar{\Delta})$ is the Cooper-pair localization length for the ES hopping. In the expression of the localization length,  $\bar{g}$ and $\bar{E}\sim E_{c}$ are the typical values of the conductance and Coulomb correlation energy between neighboring grains, respectively. The theoretical predictions of Eq.~(\ref{Eq-ES-hopping}) are least-squares fitted to our experimental data, and the results are plotted as solid lines in Fig.~\ref{FigR-H1}(d). From Fig.~\ref{FigR-H1}(d), one can see the linear parts of the curves overlap with the theoretical predictions of Eq.~(\ref{Eq-ES-hopping}). In the fitting process, setting $b=1$, one can readily obtain the values of $\xi^{\rm CPH}$, which are 1.7 and 0.9\,nm for the $x\simeq 0.50$ and 0.52 films, respectively. According to Lopatin and Vinokur,  $\xi^{\rm CPH}$ should be smaller than the mean grain size $a$ in the insulating state, and be approximately equal to $a$ when the film turns to a superconducting state~\cite{no34}. Thus, the value of  $\xi^{\rm CPH}$ for each film is reasonable. The result that the $\rho$-$T$ curves obey the ES-type-VRH law further suggests that the low-temperature charge transport processes are governed by the Cooper pair hopping for the $x\simeq0.52$ and 0.50 films.

Inspection of Fig.~\ref{FigR-H1}(d) indicates that the $\ln\rho$ vs $T^{-1/2}$ curves deviate from the ES law with further decreasing temperature below $\sim$2.7\,K. During the $\rho$-$T$ measurements, the applied current keeps at 10\,nA, around which no electroresistance effect is observed [see the inset of Fig.~\ref{FigR-H1}(d)]. The deviation at low temperatures is, therefore, irrelevant to a negative electroresistance effect~\cite{no51}. According to Delsing \emph{et al}.~\cite{no26} and Hen \emph{et al}.~\cite{no52}, the quantum fluctuations of charge would become important at low temperatures and cause the low temperature resistance to deviate from the hopping conduction law, resulting in a finite conductance. Thus, in the Pb$_{x}$(TiO$_2$)$_{1-x}$ ($x\simeq 0.50$ and 0.52) films the resistivity could be the comprehensive effect of the ES-hopping and quantum fluctuations of Cooper pairs. We note in passing that the behaviors of $\rho$ vs $\mu_0 H$ and $\rho$ vs $T^{-1/2}$ for the $x\simeq 0.55$ film are similar to those for the $x\simeq 0.52$ film.

\begin{figure}[htp]
\begin{center}
\includegraphics[scale=1.08]{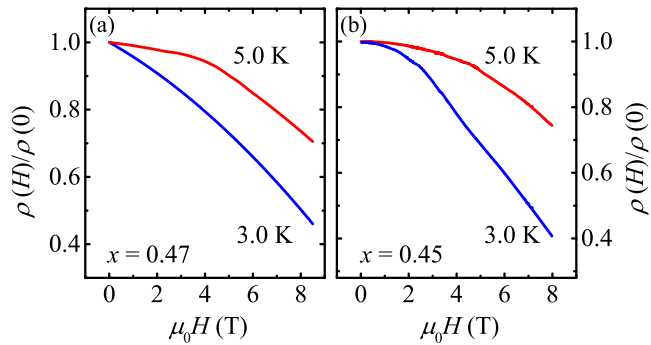}
\caption{\label{FiGLowx}The normalized resistivity $\rho(H)/\rho(0)$ vs magnetic field at 3.0 and 5.0\,K for (a) $x\simeq 0.47$ and (b) $x\simeq0.45$ Pb$_{x}$(TiO$_2$)$_{1-x}$ films.}
\end{center}
\end{figure}

Figure~\ref{FiGLowx}(a) and \ref{FiGLowx}(b) show the low-temperature magnetoresistances of the $x\simeq 0.47$ and 0.45 Pb$_{x}$(TiO$_2$)$_{1-x}$ films, respectively. Interestingly, the low-temperature magnetoresistances of the two films are all negative, indicating that single-electron hopping dominates the hopping transport. In the insulating Pb$_{x}$(TiO$_2$)$_{1-x}$ granular films, the average separation between two adjacent Pb granules increases with decreasing $x$, which will greatly enhance the normal-state intergrain resistance $R_N$. On the other hand, the Josephson energy $E_J$ between neighbor grains is $E_J = R_Q \Delta/2R_N$, where $R_Q = h/4e^2$ is the quantum resistance of Cooper pair~\cite{no21}. For a 3D periodic cubic granular array, the intergrain resistance $R_N$ can be estimated via $R_N\simeq \rho_N/a$ with $\rho_N$ being the normal-state resistivity.  Taking the normal-state resistivity $\rho_N$  as the resistivity at 15\,K, we obtained the average $E_J$ between neighbor grains for the $x\lesssim 0.52$ films and listed them in Tabe~\ref{TableI}. From Tabe~\ref{TableI}, one can see that the Josephson energies of the 0.47 and 0.45 films are reduced by $\sim$2 orders of magnitude compared to that of the $x\simeq 0.50$ film. In this situation, the intergrain Josephson coupling becomes too weak for Cooper pairs to hop between the adjacent superconducting Pb granules. As a result, the electrical transport process undergoes a transition from a Cooper-pair-dominated hopping to a single-electron-dominated hopping as $x$ decreases from 0.50 to 0.47.

\begin{figure}[htp]
\begin{center}
\includegraphics[scale=0.8]{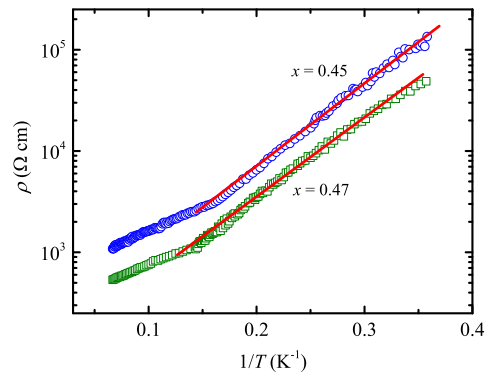}
\caption{\label{FiGLowxR-T} Resistivity  vs $T^{-1}$ from 15 down to 2.8\,K at zero magnetic field for the $x\simeq 0.47$ and 0.47 Pb$_{x}$(TiO$_2$)$_{1-x}$ films.}
\end{center}
\end{figure}

Figure~\ref{FiGLowxR-T} shows the zero-field resistivity (in a logarithmic scale) as a function of $T^{-1}$ for the $x\simeq 0.47$ and 0.45 Pb$_{x}$(TiO$_2$)$_{1-x}$ films, as indicated. In this figure, $\log_{10}\rho$ ($\ln\rho$) variation with $T^{-1}$ can be treated as a straight line from $\sim$5.0 down to $\sim$2.8\,K for each film. For single-electron hopping, the concentration of single-electron excitations follows $n\propto\exp(-\Delta/k_B T)$. Thus, the temperature dependence of resistivity obeys~\cite{no5}
\begin{equation}\label{Eq-EHP}
 \rho(T)=\rho_{02} \exp\left(\frac{\Delta}{k_B T} \right),
\end{equation}
where $\rho_{02}$ is a prefactor. We compare the experimental $\rho(T)$ data with Eq.~(\ref{Eq-EHP}), and the theoretical predictions of Eq.~(\ref{Eq-EHP}) are presented in Fig.~\ref{FiGLowxR-T} by solid lines. For the $x\simeq 0.47$ (0.45) Pb$_{x}$(TiO$_2$)$_{1-x}$ film, the fitted value of $\Delta$ is 1.55 (1.62)\,meV, which is close to the zero-temperature superconducting gap $\Delta_0$ (1.37\,meV) of Pb obtained via single-electron tunneling spectra measurements. Thus, the fact that the low-temperature $\rho$-$T$ data can be well described by Eq.~(\ref{Eq-EHP}) further confirms that single-electron hopping dominates the low-temperature transport process in the $x\simeq 0.47$ and 0.45 films. For the two films, the average charging energy of single Pb granule is still less than the superconducting energy gap of Pb. Our results therefore suggest that weak Josephson coupling between two neighbor grains is necessary for Cooper pair hopping. When the average Josephson energy between neighbor grains is too small, Cooper-pair-hopping-dominated transport cannot be observed even if the condition $E_c < \Delta$ is satisfied.

Before finishing up, we compare the superconductor to insulator transition (SIT) of the 3D Pb$_{x}$(TiO$_2$)$_{1-x}$ films with that observed in other material systems. From a macro point of view, the overall properties of the 3D Pb$_{x}$(TiO$_2$)$_{1-x}$ films are consistent with those observed in other granular systems, including 3D Al-AlO$_x$ films~\cite{no53}, 3D Al-Ge films~\cite{no36,add6}, 3D Pb-SiO$_2$ films~\cite{no38,add7}, 2D In-InO$_x$ films~\cite{no40}, and quench-evaporated 2D granular films (ultrathin films of Sn, Pb, Ga, Al, and In)~\cite{no2,add1,add2}. The commonalities of these systems are that they all experience global superconductivity, transition, and insulating regions. In the global superconductivity region, the grains are strongly coupled and the system can be treated as a dirty type-II superconductor. In this region, there are no significant differences in properties between the Pb$_{x}$(TiO$_2$)$_{1-x}$ films and other granular superconductor systems. In the transition region, most granular superconductor systems would undergo an intermediate anomalous metallic state and then reveal reentrant behavior before entering the insulating region~\cite{no2,add1,add2,add4,no38,add7}. In the intermediate metallic state, the resistance of the sample drops several decades as temperature is lowered and tends to saturate at a constant with further decreasing temperature. For the 3D Pb$_{x}$(TiO$_2$)$_{1-x}$ series, the intermediate metallic state presents in the $x\simeq 0.56$ film and the reentrant behavior occurs in the $x\simeq 0.55$ and 0.52 films as mentioned above. While for the 3D Pb$_{x}$(SiO$_2$)$_{1-x}$ system, the intermediate metallic state also appears in the films with $x$ near $\sim$0.57, and the reentrant behavior occurs in the $0.50 < x < 0.57$ films~\cite{no38}. For the 3D Pb$_{x}$(TiO$_2$)$_{1-x}$ system, the SIT occurs near $x\simeq 0.55$ and 0.56. Thus, the average critical junction resistance for SIT is $\rho_N/a\sim 5.8\times 10^6$\,$\Omega$, which is far larger than the theoretical value (9.7\,k$\Omega$)~\cite{no54} and about $\sim$4 times as large as that for the Pb$_{x}$(SiO$_2$)$_{1-x}$ series~\cite{add7}. In the insulating region, hopping conductance is prevalent and the temperature dependence of the resistance of the samples generally exhibits a significant increase in resistance below the local superconducting transition temperature $T_c^L$ due to opening an extra gap $\Delta$~\cite{no5,add8}. As mentioned in section~\ref{SecI}, the single-electron hopping prevails in the insulator region of most granular superconductors. In the Pb$_{x}$(TiO$_2$)$_{1-x}$ films with $x\simeq 0.52$ and 0.50, we observed the ES-type-VRH Cooper-pair-hopping by using the high-$k$ dielectric TiO$_2$ as the insulating matrix.

\section{Conclusion}
In summary, we successfully realized Cooper-pair hopping in insulating Pb$_{x}$(TiO$_2$)$_{1-x}$ ($x\simeq 0.50$, 0.52, and 0.55) granular films by using the high-$k$ dielectric TiO$_2$ as the insulating matrix. For these films, the low-temperature  magnetoresistances are positive under low field, and the temperature dependent behavior of resistivity obeys the ES-type VRH law in low-temperature region. As $x$ is reduced from $\sim$0.50 to $\sim$0.47, a crossover from Cooper-pair-dominated hopping to single-electron-dominated hopping is observed. These $x \lesssim 0.47$ films possess negative magnetoresistance at low temperatures and the resistivity vs temperature obeys a law of thermal-activation form. For the $x \lesssim 0.47$ Pb$_{x}$(TiO$_2$)$_{1-x}$ films, the intergrain Josephson coupling is too weak for Cooper pairs to hop between the adjacent superconducting Pb granules, and the low-temperature transport can only be mediated by single-electron hopping.

\section*{Data availability statement}
All data that support the findings of this study are included within the article (and any supplementary files).

\begin{acknowledgments}
The authors are grateful to Prof. Juhn-Jong Lin for valuable discussion. This work is supported by the National Natural Science Foundation of China through Grant No. 12174282.
\end{acknowledgments}


\begin{thebibliography}{00}
\bibitem{note1} For the convenience of discussion, we use conductivity instead of resistivity here.
\end{thebibliography}
\end{document}